\documentclass[aps,prl,reprint,groupedaddress,amsmath,floatfix,amssymb]{revtex4-1}
\usepackage{graphicx}  
\usepackage{dcolumn}   
\usepackage{bm}        
\usepackage{verbatim}   
\usepackage{braket}
\usepackage[breaklinks=true]{hyperref} 

\begin{document}

\title{Stochastic Feshbach Projection for the Dynamics of Open Quantum Systems}
\author{Valentin Link}
\affiliation{
   Institut f\"ur Theoretische Physik, Technische Universit\"at Dresden, D-01062 Dresden, Germany
   }
\author{Walter T. Strunz}
\affiliation{
   Institut f\"ur Theoretische Physik, Technische Universit\"at Dresden, D-01062 Dresden, Germany
   }

\date{\today}
\begin{abstract}
We present a stochastic projection formalism for the description of quantum dynamics in Bosonic or spin environments.
The Schr\"odinger equation in coherent state representation with respect to the environmental
degrees of freedom can be reformulated by employing the Feshbach partitioning technique for open quantum systems 
based on the introduction of suitable non-Hermitian projection operators. In this picture the reduced state of the 
system can be obtained as a stochastic average over pure state trajectories. The corresponding non-Markovian 
stochastic Schr\"odinger equations include a memory integral over the past states. 
In the case of harmonic environments and linear coupling the approach gives a new form of the 
established non-Markovian quantum state diffusion (NMQSD) stochastic Schr\"odinger equation without
functional derivatives. Utilizing spin coherent states, the evolution equation for spin environments resembles
the Bosonic case with, however, a non-Gaussian average for the reduced density operator.
\end{abstract}

\maketitle

\textit{Introduction.}-- Fueled by novel applications in the field of quantum information \cite{devega17,amin09,*costa16,*bylicka14}, 
there has been a long-lasting and growing interest in the description of open quantum system dynamics during 
the last decades \cite{breuer16}. 
The celebrated Gorini-Kossakowski-Sudarshan-Lindblad equation \cite{Lindblad76,*gorini76} is a popular tool, 
mainly due to its simple
mathematical structure \cite{breuer02,*weiss}. However, for this equation to hold one needs to make severe 
assumptions on the interplay between system and environment, in particular the \textit{Markov}, or memoryless, 
approximation. This approximation loses its validity when structured environments or strong system-environment 
couplings are considered.

Describing \textit{non-Markovian} open quantum system dynamics is a much more challenging task that inspired the 
development of several quite distinct theoretical methods \cite{devega17,breuer02,*weiss}. Arguably the most general 
form of open system evolution equation is the famous Nakajima-Zwanzig master equation \cite{nakajima58,*zwanzig60}. 
This equation includes memory-effects in the master equation for the reduced density operator $\rho_t$ in the form of 
an explicit time nonlocal term
\begin{equation}
 \partial_t\rho_t=\mathcal{L}_t\rho_t+\int_0^t\text{d}s\,\mathcal{K}_{t,s}\rho_s\,.
 \label{eq:NZ_master}
\end{equation}
Here $\mathcal{K}_{t,s}$ is the so-called memory kernel. This equation can be derived microscopically by employing a 
projection operator technique in the space of Hilbert-space operators. Specifically, one chooses a projector that maps 
the total state of system and environment onto the reduced system density operator. One is also able to construct a 
time-convolutionless (TCL) master equation within this formalism \cite{breuer02,shibata77,breuer99}. 
These TCL equations have explicitly time dependent (but time local) generators which are easier to handle
than equation (\ref{eq:NZ_master}) with an explicit memory integral. However, non-Markovian TCL master equations 
may not be well-defined in some cases because they require 
the existence of a certain operator inverse \cite{chru10,smirne10}. 

The structure of the Nakajima-Zwanzig master equation is very intricate, especially because starting from the form 
(\ref{eq:NZ_master}), one does not know explicit, general conditions for the memory kernel which ensures 
that the equation preserves 
density operators \cite{wilkie09,*barnett01}. Recent progress in this topic using special examples can be found in \cite{chruscinski16,*wudarski15}. Approximating the kernel with perturbation theory leads, in general, to loss of 
positivity \cite{benatti05, *breuer04}. 

A possible cure for positivity violation are quantum trajectories. The idea is to express the reduced density 
operator as an average over an ensemble of pure states
\begin{equation}
 \rho_t=\mathcal{M}\big(\ket{\psi_t(z^*)}\bra{\psi_t(z^*)}\big)\,,
 \label{eq:unraveling}
\end{equation}
where $\mathcal{M}(...)$ denotes the average with respect to a stochastic variable $z$
such that positivity is guaranteed by construction. A prominent example in the non-Markovian regime is
the \textit{non-Markovian quantum state diffusion} (NMQSD) formalism as introduced in \cite{diosi97}. This theory
describes quantum dynamics in environments consisting of Bosonic, harmonic baths with linear coupling to the system.
The NMQSD evolution equation for the stochastic pure states $\ket{\psi_t(z^*)}$ may be quite difficult to solve, 
since it involves derivatives with respect to the stochastic variable $z$. Nevertheless, the big advantage is that 
approximations for the stochastic pure states still lead to a reduced density operator $\rho_t$ that is positive 
by construction, see (\ref{eq:unraveling}). Apart from their numerical efficiency in applications 
\cite{roden09,suess14,hartmann17}, NMQSD and related equations also appear in exactly soluble open quantum
system models \cite{jing10}, in
non-Markovian generalizations of spontaneous wave function collapse \cite{bassi09}, in general
Gaussian open quantum system dynamics \cite{ferialdi12,diosi14}, or in attempts to establish a non-Markovian 
continuous measurement theory \cite{diosi08,wiseman08,shabani14}.

The aim of this letter is to connect ideas from both, the projection-operator method and the non-Markovian 
quantum state diffusion formalism thus laying the foundations for a very general non-Markovian quantum trajectory
theory well beyond the current status. A Feshbach projection in connection with NMQSD is also used in 
reference \cite{jing12}, where it is applied within the NMQSD formalism. Our results here are very different 
and more fundamental as we use the projection method to $\textit{derive}$ a new non-Markovian stochastic evolution
equation looking similar to (\ref{eq:NZ_master}). 
Remarkably, this new stochastic Schr\"odinger equation does not involve the problematic functional derivatives of NMQSD.
In particular we employ a {\it stochastic} variant of the Feshbach projection formalism for open quantum 
systems that was introduced in reference \cite{chru13}. In this Feshbach formalism one applies Hilbert space projection operators to the Schr\"odinger equation of the pure total state of system and environment. 
In contrast to  \cite{chru13}, we consider 
an \textit{ensemble} of \textit{non-Hermitian} projectors that naturally brings about the unraveling (\ref{eq:unraveling}).
Crucially, it applies very generally and one is no longer restricted to harmonic Bosonic baths and linear coupling
as in NMQSD.

\textit{Stochastic Feshbach projection formalism.}-- We consider an arbitrary quantum system interacting with an environment consisting of Bosonic modes that need not be 
harmonic. The total Hilbert space of system and environment is then the product space of the respective subspaces 
$\mathcal{H}=\mathcal{H}_S\otimes\mathcal{H}_E$. In the environment Hilbert space we can define creation- and 
annihilation operators $b_i^\dagger$, $b_i$ of each mode, satisfying the commutation relation 
$[b_i,b_j^\dagger]=\delta_{ij}$. Then the \textit{Bargmann coherent states} 
$|\!\!\ket{z_1 z_2 \ldots}=|\!\!\ket{\boldsymbol{z}}$ in $\mathcal{H}_E$ are defined by 
\begin{equation}
 |\!\!\ket{\boldsymbol{z}}=\text{e}^{z_1b^\dagger_1}\otimes\text{e}^{z_2b^\dagger_2}\otimes\cdots\ket{\boldsymbol{0}}\,,
\label{eq:Bargmann}
 \end{equation}
where $\ket{\boldsymbol{0}}$ denotes the vacuum bath state $b_i\ket{\boldsymbol{0}}=0$. The double-bar ket notation is 
used to emphasize that the Bargmann states are not normalized. These states are analytical in ${\boldsymbol{z}}$ and
fulfill the completeness relation
\begin{equation}
 \boldsymbol{1}_E=\int\frac{\text{d}^{2}\boldsymbol{z}}{\pi}\text{e}^{-|\boldsymbol{z}|^2}|\!\!
\ket{\boldsymbol{z}}\bra{\boldsymbol{z}}\!\!|\,,
 \label{eq:completeness}
\end{equation}
with the integral measure $\frac{\text{d}^{2}\boldsymbol{z}}{\pi}\equiv\prod_i\frac{\text{d}\text{Re}{z}_i\text{d}\text{Im}{z}_i}{\pi}$. Moreover, coherent states are never orthogonal, their scalar product is
\begin{equation}
\langle{\boldsymbol{z}}|\!|{\boldsymbol{z'}}\rangle=\prod_{i}\text{e}^{z_i^*z_i'}\,.
 \label{eq:scalar_product}
\end{equation}
Let $\ket{\Psi_t}$ be the total state of system and environment satisfying the Schr\"odinger equation
\begin{equation}
 \text{i}\partial_t\ket{\Psi_t}=H(t)\ket{\Psi_t}\,,
 \label{eq:schroedinger}
\end{equation}
with some completely general Hamiltonian $H(t)$. As initial condition, for most of the following, we assume a zero 
temperature bath such that $\ket{\Psi_0}=\ket{\psi_0}\ket{\boldsymbol{0}}$. 
Now we switch to the coherent state representation with respect to $\mathcal{H}_E$, i.e. we consider the states 
$\braket{\boldsymbol{z}|\!|\Psi_t}\equiv \ket{\psi_t(\boldsymbol{z}^*)}$ in $\mathcal{H}_S$. The Schr\"odinger equation 
in this representation would include derivatives with respect to $\boldsymbol{z}^*$.
 
In a quite different spirit we derive a new form of this equation, by employing the Feshbach technique. The non-Hermitian 
operator
\begin{equation}
 P_{\boldsymbol{z}^*}=\boldsymbol{1}_S\otimes \ket{\boldsymbol{0}}\bra{\boldsymbol{z}}\!\!|
 \label{eq:Pz}
\end{equation}
is a projector, $P_{\boldsymbol{z}^*}^2=P_{\boldsymbol{z}^*}$, because the overlap of any unnormalized coherent state 
with the vacuum is one $\braket{\boldsymbol{z}|\!|\boldsymbol{0}}=1$ according to (\ref{eq:scalar_product}). As a 
consequence, $Q_{\boldsymbol{z}^*}=\boldsymbol{1}-P_{\boldsymbol{z}^*}$ is a corresponding orthogonal projector. 
Crucially, the initial state of the form $\ket{\Psi_0}=\ket{\psi_0}\ket{\boldsymbol{0}}$ lies in the subspace spanned 
by $P_{\boldsymbol{z}^*}$, so that $Q_{\boldsymbol{z}^*}\ket{\Psi_0}=0$. Note that $P_{\boldsymbol{z}^*}$ maps any state 
in $\mathcal{H}$ to the corresponding state in coherent-state representation evaluated at ${\boldsymbol{z}^*}$. 
Thus, the Feshbach method for this particular projector results in a closed evolution equation for 
$\ket{\psi_t(\boldsymbol{z}^*)}$,
\begin{equation}
\begin{split}
\partial_t\ket{\psi_t({\boldsymbol{z}^{*}})} =&-\text{i}\braket{\boldsymbol{z}|\!|H(t)|\boldsymbol{0}}
\ket{\psi_t({\boldsymbol{z}^{*}})}\\&-\int_{0}^{t}\text{d}s\,K_{t,s}(\boldsymbol{z}^{*})
\ket{\psi_s({\boldsymbol{z}^{*}})}\, .
\end{split}
\label{eq:TCevol}
\end{equation}
Remarkably, the only assumption that was necessary to derive this equation is that the initial state of the environment 
is the vacuum. Otherwise an additional inhomogeneous term would need to be included. The kernel operator 
(an operator in $\mathcal{H}_S$) is given by 
$K_{t,s}=\bra{\boldsymbol{z}}\!\!|H(t) W_{t,s}(\boldsymbol{z}^{*})Q_{\boldsymbol{z}^{*}}H(s)\ket{\boldsymbol{0}}$, with 
the non-unitary evolution operator
\begin{equation}
 W_{t,s}(\boldsymbol{z}^{*})=\mathcal{T} \,\text{exp}\Big(-\text{i}\int_{s}^{t}\text{d}v\,Q_{\boldsymbol{z}^{*}}H(v)\Big)\,
\end{equation}
and $\mathcal{T}$ denotes chronological operator ordering. The difficulty of the problem now obviously 
lies in computing the kernel operator. 

Since solving (\ref{eq:TCevol}) for all coherent state 
labels $\boldsymbol{z}^*$ is equivalent to 
solving the total Schr\"odinger equation (\ref{eq:schroedinger}), we can easily derive a formula for 
the reduced system 
density operator $\rho_t$. As in NMQSD, by virtue of the completeness relation (\ref{eq:completeness}), the latter can be obtained as an ensemble average of pure states according to
\begin{equation}
 \rho_t=\int\frac{\text{d}^{2}\boldsymbol{z}}{\pi}\text{e}^{-|\boldsymbol{z}|^2}
\ket{\psi_t(\boldsymbol{z}^*)}\bra{\psi_t(\boldsymbol{z}^*)}\,.
\end{equation}
This is nothing but an unraveling of the form (\ref{eq:unraveling}), when we consider the coherent state labels as 
Gaussian random variables. Most remarkably, with the Feshbach technique and the choice of non-Hermitian projectors we obtain closed 
pure state evolution equations for each of these states. 

A few remarks about the initial condition $\ket{\Psi_0}$ are in order. In our formalism it is crucial to have 
$Q_{\boldsymbol{z}^*}\ket{\Psi_0}=0$ for obtaining homogeneous equations like (\ref{eq:TCevol}). This is automatically
achieved in the zero-temperature case, but can also be assured for initial states of the form 
$\ket{\Psi_0}=\ket{\psi_0}|\!\!\ket{\boldsymbol{\xi}}$, where $|\!\!\ket{\boldsymbol{\xi}}$ is a coherent state of the environment. 
Either one performs a unitary displacement or one considers the modified projector
\begin{equation}
 P_{\boldsymbol{z}^*}= \boldsymbol{1}_S\otimes \frac{|\!\!\ket{\boldsymbol{\xi}}
\bra{\boldsymbol{z}}\!\!|}{\braket{\boldsymbol{z}|\!|\boldsymbol{\xi}}}\,.
\end{equation}
This is a well-defined object because the overlap of two coherent states is never zero. Starting from the coherent-state 
initial condition one can also realize a thermal environment by taking a Gaussian mean over the coherent state labels 
$\boldsymbol{\xi}$. It might also be possible to consider more general mixed initial states by applying the formalism 
to 'amplitudes' of the total mixed state rather than pure states, similar to what is presented in reference \cite{chru13}. 
For the sake of simplicity, we assume the vacuum bath initial condition in the following.

\textit{Spin environments.}-- It is possible to generalize the new projection method in order to establish an exact quantum trajectory 
theory for spin environments (with finite spin quantum number $S$ of a single spin). 
To this aim we introduce spin coherent states analogous to the Bargmann coherent states (\ref{eq:Bargmann}) for
Bosons. Let the ground state of 
the spin system be the state that satisfies $S_z\ket{0}=S\ket{0}$, where $S_z$ denotes the $z$-component of the spin operator.
The (scaled) ladder operator $b^\dagger$ of the spin $S$ is given by \cite{radcliffe71}
\begin{equation}
 b^\dagger=\frac{1}{\sqrt{2S}}(S_x-\text{i}S_y)=\frac{J_-}{{\sqrt{2S}}}\,,\qquad [b,b^\dagger]=\frac{S_z}{S}\,.
\end{equation}
Here, contrary to the literature, the operator is defined in such a way that the analogy to the
Bosonic environment is obvious. 
The unnormalized spin coherent state is $|\!\!\ket{z}=\text{exp}(b^\dagger z)\ket{0}$, and the overlap of two such states 
becomes 
\begin{equation}
 \braket{z|\!|z'}=\Big(1+\frac{z^*z'}{2S}\Big)^{2S}\,,
 \label{eq:spinoverlap}
\end{equation}
leading to the Bosonic results in the limit $S \rightarrow \infty$.
Note that this implies $\braket{z|\!|0}=1$ and thus $\ket{0}\bra{z}\!\!|$ is again a projector that can be used in (\ref{eq:Pz}). Crucially, 
there is also a completeness relation of the form 
\begin{equation}
 \boldsymbol{1}=\int\frac{\text{d}^2z}{\pi}m(|z|^2)|\!\!\ket{z}\bra{z}\!\!|\,.
 \label{eq:spincompleteness}
\end{equation}
In order for this relation to hold, the first $2S$ moments of the weight function $m$ must satisfy
\begin{equation}
 \int_0^\infty \text{d}\sigma \sigma^p m(\sigma)=\frac{p!(2S-p)!(2S)^p}{2S!}\,,
\end{equation}
with $p=0,...,2S$ and the natural choice is $ m(|z|^2)=\frac{2S+1}{2S}\big(1+\frac{|z|^2}{2S}\big)^{-(2S+2)}\,
$, reflecting the 
overlap (\ref{eq:spinoverlap}). 
The projector $P_{z^*}$ and the completeness relation are the only ingredients necessary for the 
Feshbach method. Thus, all previous equations for Bosonic environments, including the evolution 
equation (\ref{eq:TCevol}), can be copied for the spin environment. It is a remarkable feature that within the 
new formalism the treatment of Bosonic and spin environments is formally identical with, however, a
{\it non-Gaussian} probability distribution of the coherent state labels ${z}$
according to (\ref{eq:spincompleteness}). While in many cases a spin environment 
can be described by an effective harmonic bath, this no longer holds true beyond linear response 
approximation \cite{makri99,*prokof00}.

\textit{Non-Markovian quantum state diffusion, revisited.}-- As an important application we consider an environment of harmonic oscillators with frequencies $\omega_i$, coupled linearly to the system with 
coupling strengths $g_i$. In 
this model equation (\ref{eq:TCSSE}) is a novel evolution equation for \textit{non-Markovian quantum state diffusion} and 
the stochastic interpretation will become very appealing.

In interaction representation with respect to the free environment dynamics the Hamiltonian of 
this model reads
\begin{equation}
 H(t)=H_S\otimes \boldsymbol{1}_E+\sum_{i}\big(g_i \text{e}^{\text{i}\omega_i t} L\otimes b_i^\dagger+\text{h.c.}\big)\,.
\label{eq:NMQSDmodel}
\end{equation}
Here $H_S$ is the system Hamiltonian and $L$ an operator in $\mathcal{H}_S$. One also defines the zero-temperature 
bath correlation function by
\begin{equation}
 \alpha(t)=\sum_{i}|g_i|^2 \text{e}^{-\text{i}\omega_i t}\,.
\end{equation}
We briefly recapitulate the usual way to obtain the NMQSD stochastic Schr\"odinger equation \cite{diosi97}: here, 
the Schr\"odinger equation for $\ket{\psi_t(\boldsymbol{z}^*)}$ includes derivatives with respect to the coherent 
state labels $\boldsymbol{z}^*$. Introducing a complex Gaussian stochastic process via 
\begin{equation}
 z_t^*=-\text{i}\sum_{i} g_i \text{e}^{\text{i}\omega_i t} z_i^*\,,\qquad \mathcal{M}(z_tz_s^*)=\alpha(t-s)\,,
\end{equation}
in a functional picture the Schr\"odinger equation for $\ket{\psi_t(\boldsymbol{z}^*)}$ becomes the non-Markovian 
quantum state diffusion (NMQSD) stochastic Schr\"odinger equation
\begin{equation}
\begin{split}
\partial_t\ket{\psi_t({\boldsymbol{z}^{*}})} =&-\text{i}
H_S\ket{\psi_t({\boldsymbol{z}^{*}})}+L{z}^{*}_{t}\ket{\psi_t({\boldsymbol{z}^{*}})}\\&-L^{\dagger}
\int_{0}^{t}\text{d}s\,\alpha(t-s) \frac{\delta}{\delta z_s^{*}}\ket{\psi_t(\boldsymbol{z}^{*})}\,,
\label{eq:SSE}
\end{split}
\end{equation}
again assuming the vacuum environment initial condition. Then $\ket{\psi_t({\boldsymbol{z}^{*}})}$ can be interpreted 
as a functional of $z_t^*$, and $\rho_t$ is obtained according to (\ref{eq:unraveling}). The NMQSD equation implies 
that the influence of the environment on the bath is exclusively characterized by the bath correlation function. 
Of course, the appearance of the functional derivatives makes solving this equation quite involved (for an application see \cite{roden09}).  
A numerically exact treatment is possible using a hierarchy of stochastic pure states (HOPS) \cite{suess14}.  

With the new Feshbach formalism we can now derive this equation in a new form including the memory integral. In particular, 
for Hamiltonian (\ref{eq:NMQSDmodel}), the general equation (\ref{eq:TCevol}) becomes
\begin{equation}
\begin{split}
\partial_t\ket{\psi_t({\boldsymbol{z}^{*}})} =&-\text{i}
H_S\ket{\psi_t({\boldsymbol{z}^{*}})}+L{z}^{*}_{t}\ket{\psi_t({\boldsymbol{z}^{*}})}\\&-
\int_{0}^{t}\text{d}s\,K_{t,s}(\boldsymbol{z}^*)\ket{\psi_s(\boldsymbol{z}^{*})}\,,
\label{eq:TCSSE}
\end{split}
\end{equation}
with the Kernel operator $K_{t,s}(\boldsymbol{z}^*)$. This is a completely new version of the NMQSD 
evolution equation in closed form that is well defined by construction. It can be seen as the analogue of the 
Nakajima-Zwanzig equation (\ref{eq:NZ_master}) within the NMQSD formalism. Note that the memory integral term exactly 
replaces the term with the functional derivative of NMQSD. The kernel operator can now also be interpreted as a 
functional of $z_t^*$, which can be seen most easily by expanding this object in powers of the coupling strength. 
Explicitly up to third order in $g$ one has
\begin{equation}
\begin{split}
 & K_{t,s}(z^*)=\alpha(t-s)L^\dagger\text{e}^{-\text{i}H_S(t-s)}L\\
 &+\int_{s}^t\text{d}v\, z_v^{*}\,\alpha(t-s)L^\dagger
 \text{e}^{-\text{i}H_S(t-v)}L\text{e}^{-\text{i}H_S(v-s)}L+...\,.
\end{split}
 \nonumber
\end{equation}
The Born-like approximation described in \cite{chru13} corresponds to taking into account only 
the $z^*$-independent 
first term. If in addition one formally assigns $\alpha(t)=\kappa\delta(t)$ the usual \textit{Markov} stochastic 
Schr\"odinger equation is retained. In complete analogy to the calculations in \cite{shibata77,breuer99} for a 
density
operator, in the projection formalism one can also construct TCL evolution equations. In the framework of 
NMQSD 
this has become known 
as the $O$-operator substitution \cite{diosi98,*strunz99,strunz04}. In particular, one defines an operator
$O(t,s,\boldsymbol{z}^*)$ through 
$\frac{\delta}{\delta z^*_s}|{\psi_t(\boldsymbol{z}^{*})}\rangle = O(t,s,\boldsymbol{z}^*)|
{\psi_t(\boldsymbol{z}^{*})}\rangle$
in $\mathcal{H}_S$ so that with 
$\bar{O}_t(\boldsymbol{z}^*) = \int_0^t \text{d}s\, \alpha(t-s)O(t,s,\boldsymbol{z}^*)$ one can identify
\begin{equation}
 \int_{0}^{t}\text{d}s\,K_{t,s}(\boldsymbol{z}^*)\ket{\psi_s(\boldsymbol{z}^{*})}=L^\dagger 
\bar{O}_t (\boldsymbol{z}^{*})\ket{\psi_t(\boldsymbol{z}^{*})}\,,
\end{equation}
i.e. eq. ($\ref{eq:TCSSE}$) becomes time-local. While the existence of $K_{t,s}(\boldsymbol{z}^*)$ is guaranteed by 
construction, $\bar{O}_t(\boldsymbol{z}^{*})$ may not exist for certain times. 

To fill these concepts with life we apply our formalism to a Jaynes-Cummings-type model.
In particular, the model describes a two level system with Hamiltonian $H_S=\omega \sigma_+\sigma_-$ coupled to a 
bath of harmonic oscillators in rotating wave approximation $L=\sigma_-$. For this model one can easily compute the 
kernel operator $K_{t,s}$ and obtain the new evolution equation
\begin{equation}
\begin{split}
\partial_t\ket{\psi_t({\boldsymbol{z}^{*}})} =&-\text{i}\omega \sigma_+
\sigma_-\ket{\psi_t(\boldsymbol{z}^{*})}+z_t^{*}\sigma_- \ket{\psi_t(\boldsymbol{z}^{*})}\\& - 
\int_{0}^{t}\text{d}s\, 
\alpha(t-s) \sigma_+\sigma_- \ket{\psi_s(\boldsymbol{z}^{*})} \,.
\label{eq:TCSSEJC}
\end{split}
\end{equation}
We can use the new time-nonlocal form of the stochastic Schr\"odinger equation to obtain its TCL version. For that we make 
the ansatz $\bar{O}_t=\bar{f}_t \sigma_-$, which yields
\begin{equation}
\begin{split}
\partial_t\ket{\psi_t(\boldsymbol{z}^{*})}=&-\text{i}\omega \sigma_+\sigma_-
\ket{\psi_t(\boldsymbol{z}^{*})}+z_t^{*}\sigma_- 
\ket{\psi_t(\boldsymbol{z}^{*})}\\& - \bar{f}_t \sigma_+\sigma_- 
\ket{\psi_t(\boldsymbol{z}^{*})} \,.
\label{eq:TCLSSEJC}
\end{split}
\end{equation}
We multiply both (\ref{eq:TCSSEJC}) and (\ref{eq:TCLSSEJC}) from the left by $\sigma_-$.
Comparing both expressions gives
\begin{equation}
\begin{split}
 \bar{f}_t=\int_{0}^{t}\text{d}s\, \alpha(t-s) \frac{\psi_s^+}{\psi_t^+} \,,
\label{eq:psi+}
\end{split}
\end{equation}
with the complex-valued function $\psi_t^+$ satisfying the nonlocal differential equation
\begin{equation}
 \partial_t\psi_t^+=-\text{i}\omega \psi_t^+-\int_{0}^{t}\text{d}s\, \alpha(t-s) \psi_s^+\,.
\end{equation}
This result can also be obtained by using a Heisenberg operator technique, without knowledge of (\ref{eq:TCSSEJC}) 
\cite{strunz04}. Here it arises quite naturally from the structure of the new evolution equation. Note that $\bar{O}_t$ 
is ill-defined whenever $\psi_t^+=0$, whereas $K_{t,s}=\alpha(t-s)\sigma_+\sigma_-$ is always well defined. With 
the operator $\bar{O}_t$ at hand it is also easily possible to derive the corresponding TCL master equation for this model, 
see reference \cite{strunz04}.

\textit{Conclusions.}-- With the aim to describe non-Markovian quantum dynamics, we have introduced a new formalism applicable to 
quantum systems 
in environments consisting of Bosonic modes or spins. The very general formalism arises from a 
Feshbach-like 
projection operator technique based on a stochastic ensemble of non-Hermitian projectors in the environment 
Hilbert space. 
In particular, when the Bosonic or spin modes are initially in the vacuum state, the quantum state in coherent 
state-representation 
with respect to the environmental degrees of freedom satisfies the new evolution equation (\ref{eq:TCevol}). 

We applied the formalism to an environment of harmonic oscillators linearly coupled to the system. For this model we 
recover the non-Markovian quantum state diffusion (NMQSD) stochastic Schr\"odinger equation in a closed 
form (\ref{eq:TCSSE}), where the functional 
derivative with respect to the stochastic trajectory is replaced by a memory integral over the past state. 
Structurally, the projection method provides a new access to the NMQSD formalism that makes the non-Markovian 
nature of this equation much more apparent. While the 
new equation is always well defined, this is not true for the TCL version of the NMQSD stochastic Schr\"odinger 
equation arising from the $O$-operator method, as we have seen based on the simple example of the 
Jaynes-Cummings-type model. 
It should be 
stressed that for the formalism no restrictions on the total Hamiltonian are made. Thus the 'environment' could 
also describe an interacting Bosonic many-body system. There is currently a huge interest in non-equilibrium dynamics
of many-body quantum systems in the context of ultracold atomic gases (see, for instance \cite{trotzky12})
and we believe that the current work opens a door for a quantum open system point of view of these dynamics.
For applications in this field one needs to find suitable approximations 
for the memory kernel operator. Since the formula for the reduced state (\ref{eq:completeness}) has the form of a 
stochastic unraveling, an approximated kernel still leads to a positive reduced density operator, in contrast to 
the Nakajima-Zwanzig method, where approximations typically lead to loss of positivity.
By introducing quantum spin states analogous to the Bargmann coherent states one can also apply the formalism 
when the environment consists of spins. Notably, this leads to evolution equations whose form is identical to 
the ones for Bosonic environments with, however, a {\it non-Gaussian} distribution of the random 
variables ${\boldsymbol{z}^{*}}$.

\textit{Acknowledgment.}-- It is a pleasure to thank Nina Megier, Kimmo Luoma, Richard Hartmann and Alex Eisfeld
for numerous discussions related to this work.

\bibliography{stochastic_feshbach_05}

\end{document}